\documentclass[aps, pra, reprint,a4paper, amsfonts, amssymb, amsmath, reprint, showkeys, nofootinbib,superscriptaddress]{revtex4-1}% aps, prl, 
\usepackage[english]{babel}
\usepackage[utf8]{inputenc}
\usepackage{float}
\usepackage[disable]{todonotes} % * change to that to temprorarily disable all todo notes
\usepackage{soul,xcolor}
\setstcolor{red}%added by GJ

\usepackage{silence}
\usepackage{amsthm}
\usepackage{mathtools}
\usepackage{physics}
\usepackage{xcolor}
\usepackage{graphicx}
\usepackage[left=20mm,right=20mm,top=35mm,columnsep=15pt]{geometry} 
\usepackage{adjustbox}
\usepackage{placeins}
\usepackage[T1]{fontenc}
\usepackage{lipsum}
\usepackage{csquotes}
\usepackage{wrapfig}

\usepackage[pdftex, pdftitle={Article}, pdfauthor={Author}]{hyperref} % For hyperlinks in the PDF

\hypersetup{
    colorlinks = true,
    allcolors = {blue},
}

\makeatletter
\renewcommand*{\@fnsymbol}[1]{\ensuremath{\ifcase#1\or 
   \mathsection\or \mathparagraph\or \|\or **\or \dagger\dagger
   \or \ddagger\ddagger \else\@ctrerr\fi}}
\makeatother

\begin{document}

\title{\textcolor{black}{An optical Ising spin glass simulator with tuneable short range couplings}}

\author{Louis Delloye}
\author{Gianni Jacucci}
\affiliation{Laboratoire Kastler Brossel, Sorbonne Université, Ecole Normale Supérieure-Paris Sciences et Lettres (PSL) Research University, Centre National de la Recherche Scientifique (CNRS) UMR 8552, Collège de France, 24 rue Lhomond, 75005 Paris, France}

\author{Raj Pandya}
\affiliation{Laboratoire Kastler Brossel, Sorbonne Université, Ecole Normale Supérieure-Paris Sciences et Lettres (PSL) Research University, Centre National de la Recherche Scientifique (CNRS) UMR 8552, Collège de France, 24 rue Lhomond, 75005 Paris, France}
\affiliation{Cavendish Laboratory, University of Cambridge, JJ Thomson Avenue, Cambridge, CB3 0HE, UK}

\author{Davide Pierangeli}
\affiliation{Institute for Complex System, National Research Council (ISC-CNR), 00185 Rome, Italy}
\affiliation{Physics Department, Sapienza University of Rome, 00185 Rome, Italy}

\author{Claudio Conti}
\affiliation{Physics Department, Sapienza University of Rome, 00185 Rome, Italy}

\author{Sylvain Gigan}
\affiliation{Laboratoire Kastler Brossel, Sorbonne Université, Ecole Normale Supérieure-Paris Sciences et Lettres (PSL) Research University, Centre National de la Recherche Scientifique (CNRS) UMR 8552, Collège de France, 24 rue Lhomond, 75005 Paris, France}

%\date{\today} % Leave empty to omit a date
\begin{abstract}
    \textcolor{black}{Non-deterministic polynomial-time (NP) problems are ubiquitous in almost every field of study. Recently, all-optical approaches have been explored for solving classic NP problems based on the spin-glass Ising Hamiltonian.} However, obtaining programmable spin-couplings in large-scale optical Ising simulators, on the other hand, remains challenging. \textcolor{black}{Here, we} demonstrate control of the interaction length between user-defined parts of a fully-connected Ising system\textcolor{black}{. This is achieved }by exploiting the knowledge of the transmission matrix of a random medium and by using diffusers of various thickness. Finally, we exploit our spin-coupling control to \textcolor{black}{observe} replica-to-replica fluctuations and its \textcolor{black}{analogy} to standard replica symmetry breaking.
\end{abstract}

\maketitle

\listoftodos

\section{Introduction}
\todo[color=red]{update abstract+intro to fit the flow inversion done in last section}
% Relevance of NP-pbl
Non-deterministic polynomial-time problems (NP-problems) are important in many fields of study from \textcolor{black}{the} physical to social sciences \cite{parisi-spin, Takeda_2018, PhysRevB.99.104509, RevModPhys.80.1061,goto_10.1126_sciadv.aav2372,TSP10.1057_jors.1975.151,Ising-NP}. However, they often have intractable solve-times with classical computers as their solve-time scales exponentially with the size of the input \cite{np-pbl}. An archetypal NP-problem is finding the ground-state of an Ising spin-system \cite{Barahona_1982, Bachas_1984, Nishimori2001, Anderson, original_SK}. This is of particular interest as many wide-spread NP-problems can be analytically mapped \cite{Ising-NP,tune-spin-us} onto an Ising Hamiltonian (c.f. \autoref{eq_I_H}). Solving any \textcolor{black}{of these }particular NP problem thus reduces to finding the ground state of the corresponding Ising system.\par

% Optical Ising simulators

Recently, Ising models have been experimentally simulated in a number of ways, both by using classical and quantum systems \cite{Pierangeli:19, Pierangeli:20, Pierangeli:20-3,rand-laser-1, rand-laser-2, Johnson2011,Boixo2014, Berloff2017,Kalinin2020, Harris2018,Hamerly2019, non-lin-wave-prop, tune-spin-us, fpga-qa, Yamamoto-2,Haribara_2016,McMahon2016,Inagaki2016,Bohm2019,PhysRevApplied.13.054059,Honjo2021,davidson:2013,yamamoto:2020,strinati:2021, davidson:2020-1,davidson:2021-2, Yamamoto-1}. \textcolor{black}{A class of very }promising systems are optical Ising simulators based on optical parametric oscillators (OPOs) \cite{yamamoto:2020,Yamamoto-1,Yamamoto-2,Haribara_2016,McMahon2016,Inagaki2016,Bohm2019,PhysRevApplied.13.054059,Honjo2021,davidson:2013,yamamoto:2020,strinati:2021, davidson:2020-1,davidson:2021-2,100000-opo, Cen2022} and photonic annealers based on wavefront shaping \cite{sunQuadraturePhotonicSpatial2022a,kumarLargescaleIsingEmulation2020a, kumarObservationDistinctPhase2023, huangAntiferromagneticSpatialPhotonic2021a, fangExperimentalObservationPhase2021a}. \textcolor{black}{The latter uses propagation in complex \textcolor{black}{media} whereas the other exploits OPOs and time-multiplexing}. OPO-based methods have a high tunability but lack scalability. On the other hand, WS-based methods present a high \textcolor{black}{scalability and connectivity but are limited }in their tunability. The degree of customization of WS-based simulators can be increased by leveraging the knowledge of the transmission matrix \cite{tune-spin-us}. \par
% What we propose
In this \textcolor{black}{Letter}, we demonstrate the control of the spin interaction length by \textcolor{black}{using thin diffusive media}. By tuning the interaction \textcolor{black}{length we observe }replica-to-replica fluctuations analogous to replica symmetry breaking (RSB) in spin glasses \cite{mezard:jpa-00209816, mezardSpinGlassTheory1986}. Our results represent a relevant step toward the realization of fully-programmable Ising machines on free-sapce optical-platforms, capable of solving complex spin-glass Hamiltonians on a large scale. \par

% \section{Results}
% Figure 1
\begin{figure}[t]
    \centering
    \includegraphics[width=0.9\linewidth, page=2]{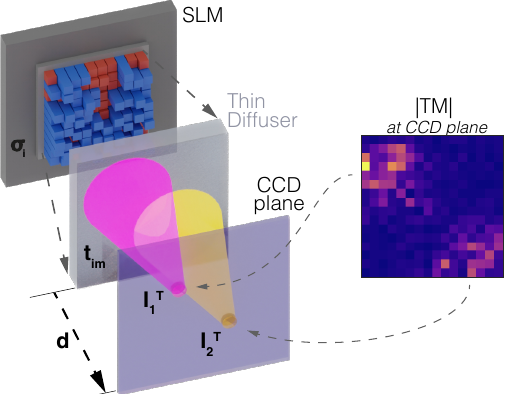}
    \caption{\textbf{Optical spin-glass simulator.} \textcolor{black}{A laser is collimated on a spatial light modulator (SLM)} and passes through a thin diffuser to then reach the imaging plane. The surface of the SLM is directly imaged (via a 4f-system) on the surface of the diffuser\textcolor{black}{. The light is focused on two output modes. }The inset shows the two rows of the transmission matrix $(t_{im})_{im}$ summed and then reshaped to SLM dimensions. The light reaching the pink (resp. yellow) focus can only come from certain pixels within the pink (resp. yellow) cone on the SLM plane ($\sigma_i$), of which the extent is given by the distance $d$ and/or the diffuser's matrix $t_{im}$\textcolor{black}{, corresponding to solving an Ising Hamiltonian with local couplings.}}
    \label{fig1}
\end{figure}
\todo{update Fig1 title}

% Figure 2
\begin{figure}[t] % 2 col
    \centering
    \includegraphics[width=0.9\linewidth, page=4]{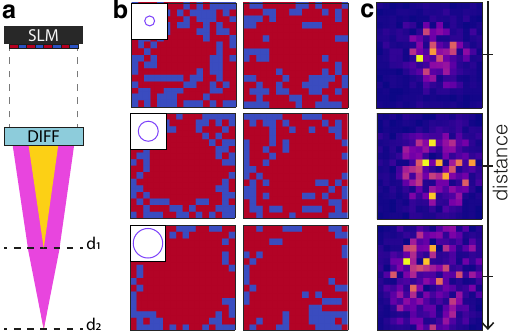}
    \caption{\textbf{Tuning the spin correlation length by optical propagation. a)} Schematic representation of the setup in two different distance configurations: $d_1$ in red and $d_2$ in orange, where the colored cones represent where the light reaching one CCD pixel comes from on the SLM. \textbf{b)} Simulation (first col) and experimental (second col) spin maps at three different imaging plane CCD-DIFF distances. The further \textcolor{black}{away} the plane, the \textcolor{black}{wider} the correlation. \textbf{c)} The amplitudes of the experimental transmission matrices at these same three distances.}
    \label{fig2}
\end{figure} % 2 col

%% Def Ising systems + description setup
\textcolor{black}{A system of $N$ coupled spins is described by the Hamiltonian in \autoref{eq_I_H}. For the Sherrington-Kirkpatrick model \cite{Nishimori2001}, the couplings ($J_{ij}$) are all-to-all random couplings, and are drawn from a Gaussian distribution}. Finding the ground-state of \autoref{eq_I_H} is an NP-hard problem. This problem can be mapped onto a physical hardware thanks to coherent light-propagation in disordered medium\textcolor{black}{. In this configuration,} one can show that the transmitted intensity after a multiply-scattering medium takes the form \cite{Pierangeli:20}:
\begin{equation}
    H = - \sum_{i,j=1}^N J_{ij} \sigma_i \sigma_j = - I
    \label{eq_I_H}
\end{equation}
with $J_{ij} = -\sum_m^M \Re{\overline{t_{im}} t_{jm}}$, where $m$ runs on the output modes and $t_{im}$ and $t_{jm}$ are transmission matrix ($T$) elements \textcolor{black}{of the random medium}. The matrix $T$ links the input modes on the spatial light modulator (SLM) ($E_{in}$) and the output modes on the camera ($E_{out}$) via $E_{out}=T E_{in}$ \cite{2011-Popoff}. \textcolor{black}{The spins are encoded on the input modes (pixels) with a binary phase of $0$ and $\pi$, corresponding to $\pm 1$ spin states, respectively \cite{Pierangeli:20, Leonetti:20}.} \textcolor{black}{Thanks to the physical system described in \autoref{fig1} and using an optimization algorithm one can find the ground state of the Ising problem \cite{Pierangeli:19}.}\newline

% Figure 3
\begin{figure}[ht]
    \centering 
    \includegraphics[width=0.9\linewidth, page=3]{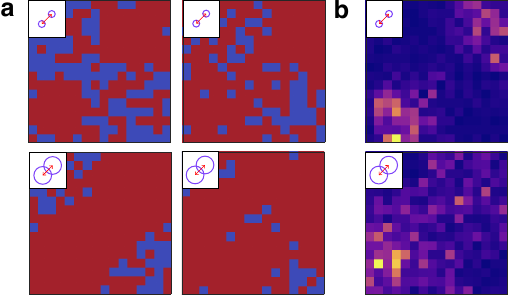}
    \caption{\textbf{Controlling the interaction between spin clusters.}  \textcolor{black}{\textbf{a)} Simulation (first columns) and experimental (second columns) spin ground states. \textbf{b)} TM amplitudes obtained summing all rows of the measured TM.} Two \textcolor{black}{isolated} regions of magnetized spins \textcolor{black}{(when optimizing over to two foci at the CCD plane) as displayed on the SLM} (first row). As the distance between CCD and diffuser increases the extent of the \textcolor{black}{magnetized regions also increases}, the regions start to overlap and therefore interact (second row).}
    \label{fig3}
\end{figure}
\todo[color=yellow]{fig 3 the red looks faded compared to fig2}

The couplings distribution ($J_{ij}$) can be tuned by using the knowledge of the transmission matrix $T$ \cite{tune-spin-us}. In the present work, we use thin diffusive media \cite{electro-optical-building,sebbah2001waves}. These allow us —in contrast to thick scattering media \cite{Pierangeli:20,Pierangeli:19,Pierangeli:20-3}— to control the extent of spins contributing to the Hamiltonian. The further the camera is from the diffuser, the more the speckle pattern will spatially expand and the more the light coming from various input modes is mixed. In terms of couplings, it means that for a given input mode, the number of connected other input modes will increase as we get further from the medium. This effect can be used to control the spatial extent of the spins' couplings. The insets in \autoref{fig2} show a graphical representation of the amount of pixels contributing to the intensity on one single CCD pixel at two different distances. In terms of spin couplings, it means that for a given spin, the number of connected spins will increase as we get further from the medium. This effect can be used to control the spatial extent of the spin interaction. Remarkably, there exists two extreme points. The first one being when the distance is such that the speckle is fully developed, leading to an all-to-all coupled spin system (e.g. at position $d_2$ in \autoref{fig2}). The second one is when the distance is such that the spins are not coupled at all (e.g. at position $d_1$ in \autoref{fig2}). In this situation, the distance is so small that the light reaching the specific output can only come from one input mode (or SLM macro pixel)\textcolor{black}{. In other words,} it is an Ising system of N decoupled spins.\newline

% Figure 4
\begin{figure*}[ht]
    \centering
    \includegraphics[width=0.9\linewidth, page=1]{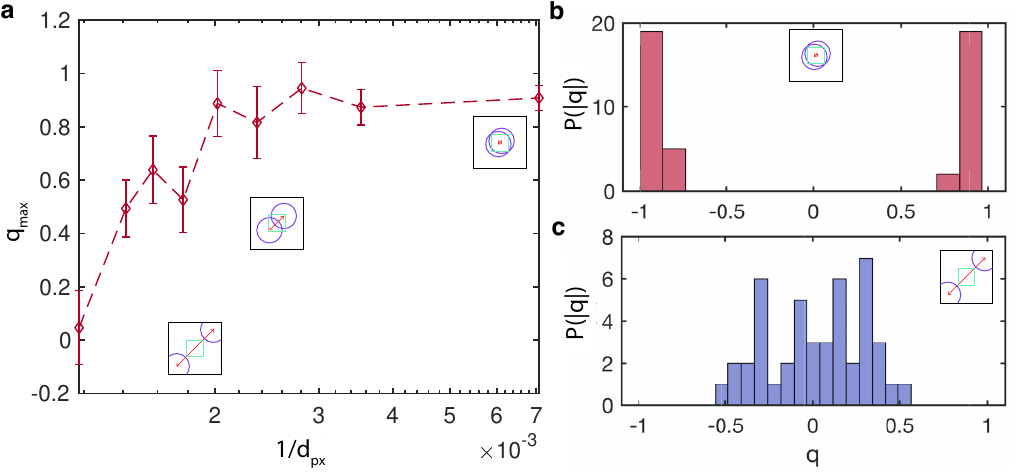}
    \caption{\textbf{Controlling frustration and replica symmetry breaking. a)} Two regions of fixed size are increasingly getting closer on the SLM plane. Each point corresponds to ten replicas for one given distance. The graph shows the maximum of the probability distribution of the correlation between replicas $q_{max}$  as a function of the distance between the regions (in px on the CCD image). Replica-to-replica fluctuations vanish when the correlation is close to 1. \textbf{b)} and \textbf{c)} Histograms of the ground-state correlation for different replicas at  maximum and minimum distance respectively.}
    \label{fig4}
\end{figure*}
% Figure 2
In Reference \cite{tune-spin-us} we demonstrate the ability to tune the Hamiltonian of the system such that we obtain magnetized ground states for a given set of couplings. As shown in \autoref{fig2}, we apply the same technique to thin diffusers to tune the couplings such that we obtain a ground state with a localized magnetization. \autoref{fig2}\textcolor{blue}{a} shows, for both simulations and experiments, three ground states for three different (increasing) distances between diffuser and CCD. \textcolor{black}{As shown in \autoref{fig2}, }we observe that the magnetized region is localized and has a finite size, whereas fully-magnetized ground states are expected for a thick medium  \cite{tune-spin-us}. \autoref{fig2}\textcolor{blue}{c} offers a preview of the amplitude of the transmission matrix coefficients ($|T|$) for the three previous distances. It is evident here that the light that reaches this given CCD pixel (chosen for the optimization) comes from only a subset of the input modes (i.e. spins). Secondly, one can note that the interaction length between the spins grows with the distance to the diffuser (c.f. \autoref{fig2}\textcolor{blue}{c}). \par

% Figure 3
The ability to tune the spatial extent of the region where spins are mutually coupled allows us to create several clusters and explore the interactions between them \textcolor{black}{by selecting two pixels at some distance, and optimizing over the sum of their intensities.} We define the interaction area as the zone where the two regions overlap. The spins within this area contribute to both foci at the two different camera-pixels corresponding respectively to the two regions. \autoref{fig3} displays an example of two regions\textcolor{black}{, varying their distance and the interaction length.} In detail, \autoref{fig3}\textcolor{blue}{a} shows the ground states, for both simulations and experiments, obtained for the whole system (both regions) and for two different distances between diffuser and CCD. The first column shows the case where the two regions are not overlapping, the second column shows the case where the two regions are overlapping. \autoref{fig3}\textcolor{blue}{b} shows the transmission matrix amplitudes for the two previous cases. We can \textcolor{black}{also} tune the overlap between spin clusters by keeping the distance $d$ fixed and by translating the two pixels on the CCD — effectively optimizing for two closer pixels.\par

% There exists several ways to tune the overlap between regions. As seen above, changing the distance between diffuser and CCD is one of them, as it will make both regions bigger and therefore the interaction region too. Another one, is to keep the distance diffuser-to-camera fixed but to \emph{translate} the two pixels on the CCD and get them closer (i.e. select two closer or further pixels) — effectively optimizing for two closer pixels. \autoref{fig4} uses the latter as we will see below.\par
\todo{keep commented section ?}

% Figure 4
We finally investigate the probability of the two interaction regions being magnetized in \textcolor{black}{an uncorrelated way} as their distance varies. \textcolor{black}{When regions are not coupled, they are independent and the magnetization is thus the same $50\%$ of the time. When they get closer, their magnetization tends to be correlated. This is due the fact that the regions are not independent anymore, as there is an overlap between them, corresponding to a coupling between the two regions.} \textcolor{black}{To quantify the correlations between various replica, we defined—in analog manner as the Parisi order parameter \cite{non-lin-wave-prop, rand-laser-2, M_Mezard_1992, parisi-mezard, parisi-order}—the following metric for fully-magnetized spin configurations:}
$$
q_{\alpha \beta} = 1 - \frac{|\sum_{i=1}^N X^\alpha_i - X^\beta_i|}{N}
$$
where $\alpha$ and $\beta$ refer to two replicas, $N$ is the number of spins in the region of interest and $X^{\alpha,\beta}_i \in \{-1,1\}$ are spin configurations. The zone of interest in which the correlation between replica ground states is calculated \textcolor{black}{is} a smaller region than the full input modes mask (c.f. green square in inset schematics of \autoref{fig4}\textcolor{blue}{a}).\par

\autoref{fig4}\textcolor{blue}{a} shows the correlation between replicas as a function of the distance between the two regions. The correlation is defined as the maximum of the probability distribution of the correlation between replicas $q_{max}$ (which is defined as $q_{\text{max}} = \text{argmax}_q P(q)$, where $P$ represents the distribution\todo[color=red]{i need to renormalize for it to be a probability (1 must be the max)} of the degree of correlation $q$ between ground states). The correlation is close to unity when replica-to-replica fluctuations drop to $0$. \autoref{fig4}\textcolor{blue}{b} and \autoref{fig4}\textcolor{blue}{c} show the histograms of the correlation between ground states of different replicas when the two regions are the furthest and the closest respectively. We observe that by increasing the overlap between the two areas —and therefore their interaction— the correlation increases, in analogy with the replica symmetry breaking transition, typical of random spin systems (c.f. \autoref{fig4}\textcolor{blue}{c}). A simulation of our system can be found in the \textcolor{blue}{Appendix C}, describing some conditions on the overlap for which \textcolor{black}{one observes} RSB-like behavior and what the assumptions on the couplings amplitude distribution \textcolor{black}{are}.\par

% Conclusion
In short, we have \textcolor{black}{observed and} demonstrated experimental control over the interaction length of an Ising spin-glass system based on free-space optics and disordered media. We have also shown that we can control the interaction between two regions of spins and induce replica symmetry breaking.

% outlooks & perspectives (talk about the future)
The proposed system is a step towards encoding more complex Hamiltonians in hope of solving more complex NP-problems. It is also a new platform for studying replica symmetry breaking. Moreover, the degeneracy changing when regions interact means that the systems seems to develop long-range couplings from short-range couplings. This effect could be leveraged as a new type of annealing approach. Indeed, one could anneal a subset of the system and then \textcolor{black}{— driven by the interaction —} the whole system would anneal.

Furthermore, one could even consider generalizing this Ising system to a Hopfield system \cite{Leonetti:20} as our experimental setup is algorithm-agnostic and the spins are defined with a continuous encoding, therefore a continuous orientation of spins could be explored. Other platforms could also be envisioned, such as using multiple SLMs or using non-linear or more complex media to obtain more complex couplings distributions (bimodal, multimodal, etc). \par

%TC:ignore
\begin{acknowledgments}
L.D., G.J., R.P., D.P., C.C., and S.G. designed the project.  L.D. carried out experiments and data analysis, G.J. and R.P. numerical simulations. L.D. and D.P. wrote the paper with contributions from all the authors. This project was funded by the European Research Council under the grant agreement No. 724473 (SMARTIES). R.P. thanks Clare College, University of Cambridge for funding via a Junior Research Fellowship. 
\end{acknowledgments}

\appendix
\section*{Methods}

{\bf Experimental setup.} 
We evaluated experimentally our approach for controlling the couplings of the spin simulator (c.f. \autoref{fig1}) as well as its extension with the additions described below. A laser (Coherent Sapphire SF 532, $\lambda = 532 nm$) is directed onto a reflective phase-only, liquid-crystal SLM (Meadowlark Optics HSP192-532, $1920 \times 1152$ pixels\textcolor{black}{, aggregated into $N=256$ macro-pixels}) divided into N macro-pixels (spins). The Fourier transform of the modulated light is projected on the objective back focal-plane (OBJ1, $10 \times$, $\mathrm{NA} = 0.1$) and focused on a scattering medium (DIFF). \textcolor{black}{As a scattering, medium we used a surface-diffuser that is commercially available (Edmund, 12.5mm, 25°). In practice, using a thin volumetric diffuser or combining a surface diffuser and free-space propagation are equivalent in our scheme.} The scattered light is then collected by a second objective (OBJ2, $20 \times$, $\mathrm{NA} = 0.4$) and the transmitted intensity is detected by a CCD camera (Basler acA2040-55$\mu$m, $2048 \times 1536$ pixels). The spins and bias (from the TM \cite{tune-spin-us}) are encoded by a spatial light modulator (SLM) in a phase pattern whose binary part is sequentially updated until the ground-state is reached. Note that for the optimization any algorithm can be used, i.e., the setup is algorithm agnostic, as the advantage of the aforementioned simulator resides in the parallel measurements of the energy \cite{Pierangeli:20}.\par

{\bf Transmission matrix calculation and ground-state search.}
The transmission matrix of the scattering medium was estimated as in \cite{2011-Popoff}. In detail, each row of the TM can be reconstructed by monitoring how the intensity on a given CCD pixel changes when a phase modulation is applied to the input patterns on the SLM.\textcolor{black}{ Those interferometry measurements provide the TM. Taking the phase conjugate of this matrix gives the SLM mask necessary for proper focusing \cite{tune-spin-us}}. The TM is sensitive to translations and rotations of the scattering medium as well as to the input and detection hardware. In this work, we define the stability\textcolor{black}{-time} as a variation within $10\%$ of it original value. This time \textcolor{black}{(typically $\sim 120$ minutes)} is long enough to run our experiments but for larger systems one would need more stable architectures. The ground-state search is conducted sequentially by means of the recurrent digital feedback. Computation starts from a random configuration of N binary macro-pixels (spins) on the SLM. The measured intensity distribution determines the feedback signal.  At each iteration, an arbitrary batch of spins is randomly flipped if it increases the intensity at a chosen output mode. The batch size decreases over the optimization procedure, starting from $12\%$ of the pixels to a single pixel for the last $\sim 600$ iterations. \par

{\bf Numerical methods.}
The numerical model used in this work is a generalization of \cite{Pierangeli:20}. The optical SG is numerically simulated by forming N pixel blocks (SLM plane). The initial optical field has a constant amplitude, and its phase is a random configuration of N binary phases, $\phi_i=0, \pi$. A \textcolor{black}{gaussian i.i.d.} transmission matrix T with random complex numbers is generated. At each iteration, a randomly selected single spin is flipped. The input phase \textcolor{black}{is updated} if the output total intensity increased after the linear propagation of the field. The bias in the numerical framework is calculated as in the experiment by starting from the knowledge of T. Numerical evaluation of $I_T$ corresponds to a measurement with a detector in a noiseless system. In general, within this scheme, $\sim 10N$ iterations are sufficient for a good convergence, i.e., when focus intensity reaches a plateau.  All codes are implemented in MATLAB on an Intel processor with 14 cores running at 3.7 GHz and supported by 64 GB ram.
\par

\todo{supplementary figures ? }
% \section*{Appendix C: Supplementary figures}
% \label{si}
% \begin{figure}[h]
%     \centering
%     \includegraphics[width=0.9\linewidth, page=1]{SI/figs/TM_3dist.pdf}
%     \label{TMs}
%     \caption{\textbf{Simulated Transmission Matrices.} The various simulated transmission matrices. Two center are choose and the distance to the diffuser is simulated by changing the chosen standard deviation of the corresponding gaussian distributions. Each row corresponds to a given standard deviation and each column are a different distance of the two chosen foci on the CCD plane.}
% \end{figure}

% \begin{figure}[h]
%     \centering
%     \includegraphics[width=0.9\linewidth, page=1]{SI/figs/overlapFdist.pdf}
%     \label{overlap}
%     \caption{\textbf{Distribution overlap as a function of distance} This figure shows the overlap (defined as a 2D integral) between the two simulated distributions as a function of the distance between the two foci. The overlap is calculated for three given standard deviation of the gaussian distributions (c.f. three colors).}
% \end{figure}

% \begin{figure}[h]
%     \centering
%     \includegraphics[width=0.9\linewidth, page=1]{SI/figs/axis_change.pdf}
%     \label{axis-change}
%     \caption{\textbf{Replica-to-replica symmetry breaking as a function of overlap} This figure shows the value of the position of the maximum of the probability distribution $q_max$ as a function of the overlap between the two distributions. The overlap is calculated for three given standard deviation of the gaussian distributions (c.f. three colors).}
% \end{figure}

\bibliography{main.bib}

\end{document}